# Design of Fault-Tolerant and Dynamically-Reconfigurable Microfluidic Biochips*

Fei Su and Krishnendu Chakrabarty

Department of Electrical & Computer Engineering

Duke University, Durham, NC 27708

E-mail: {*fs, krish*}@ee.duke.edu

**Abstract**

*Microfluidics-based biochips are soon expected to revolutionize clinical diagnosis, DNA sequencing, and other laboratory procedures involving molecular biology. Most microfluidic biochips are based on the principle of continuous fluid flow and they rely on permanently-etched microchannels, micropumps, and microvalves. We focus here on the automated design of "digital" droplet-based microfluidic biochips. In contrast to continuous-flow systems, digital microfluidics offers dynamic reconfigurability; groups of cells in a microfluidics array can be reconfigured to change their functionality during the concurrent execution of a set of bioassays. We present a simulated annealing-based technique for module placement in such biochips. The placement procedure not only addresses chip area, but it also considers fault tolerance, which allows a microfluidic module to be relocated elsewhere in the system when a single cell is detected to be faulty. Simulation results are presented for a case study involving the polymerase chain reaction.*

## 1. Introduction

Microfluidics-based biochips are receiving considerable attention nowadays [1]. These composite microsystems, which manipulate fluids at nanoliter-to-microliter scales, can greatly simplify cumbersome laboratory procedures. Such lab-on-a-chip devices are therefore expected to facilitate *in-vitro* clinical diagnosis, DNA sequencing, and other common procedures in molecular biology.

Most microfluidic biochips of today, consisting of permanently-etched micropumps, microvalves, and microchannels, are based on the principle of continuous fluid flow [1]. A more promising approach is to manipulate liquids as discrete microdroplets. This novel droplet-based technique is referred to in the literature as "digital microfluidics" [2]. Each droplet can be controlled independently and each cell in the microfluidic array has the same structure. In contrast to the continuous-flow systems, digital microfluidics offers dynamic reconfigurability as well as a scalable system architecture [3]. Groups of cells can be dynamically reconfigured to change their functionality during the execution of a bioassay. Multiple assays can be concurrently carried out on the microfluidic platform [4].

The complexity of digital microfluidics-based biochips is expected to steadily increase due to the need for multiple and concurrent assays on the chip. Time-to-market and fault tolerance are also expected to emerge as design considerations. As a result, current full-custom design techniques will not scale well for larger designs. There is a need to deliver the same level of CAD support to the biochip designer that is now available to the semiconductor industry. Moreover, it is expected that these microfluidic biochips will be integrated with microelectronic components in next-generation system-on-chip designs. The 2003 International Technology Roadmap for Semiconductors (ITRS) clearly identifies the integration of electrochemical and electro-biological techniques as one of the system-level design challenges that will be faced beyond 2009, when feature sizes shrink below 50nm [5].

Early research on CAD for digital microfluidics-based biochips has been focused on device-level physical modeling of single components [6]. While top-down system-level design tools are now commonplace in IC design, no such efforts have been reported for digital microfluidic chips. Here we propose a design methodology that attempts to apply variants of classical module placement techniques to the design of digital microfluidics-based biochips, and thus reduce design time and human effort.

We envisage the following steps in the synthesis of biochips. A behavioral model for a biochemical assay is first generated from the labotorary protocol for that assay. Next, architectural-level synthesis is used to generate a macroscopic structure of the biochip; this structure is analogous to a structural RTL model in electronic CAD. The macroscopic model provides an assignment of assay functions to biochip resources, as well as a mapping of assay functions to time-steps, based in part on the dependencies between them. Finally, geometry-level synthesis creates a physical representation at the geometrical level, i.e., the final layout of the biochip consisting of the configuration of the microfluidic array, locations of reservoirs and dispensing ports, and other geometric details.

Many biochips are expected to be used for safety-critical applications, e.g., patient health monitoring, neo-natal care, and the monitoring of environmental toxins. Therefore, these biochips must be designed to be fault-tolerant such that they can continue to operate reliably in the presence of faults. One approach to fault tolerance is to carefully include spare cells in the array such that faulty cells can be bypassed without any loss of functionality. The locations of the spare cells must be determined by the physical design tool that maps modules to sets of cells in the array.

A key problem in the design of biochips is the placement of microfluidic modules such as different types of mixers and storage units. The ability to reconfigure the microfluidic array during the execution of bioassays makes this placement problem different from the traditional placement problem in electronic design. Furthermore, the placement of the microfluidic modules has a strong impact on the ease of reconfigurability for fault tolerance. Thus, in addition to area (measured by the number of cells in the array), fault tolerance is also a placement criterion.

In this paper, we focus on the problem of module placement for digital microfluidics-based biochips with area and fault tolerance as the placement criteria. An example of a real-life biochemical procedure, i.e., polymerase chain

* This research was supported by the National Science Foundation under grant number IIS-0312352

reaction (PCR), is used to evaluate the proposed methodology. Since the placement problem is known to be NP-complete [7], a simulated annealing-based heuristic approach is developed to solve the problem in a computationally efficient manner. Solutions for the placement problem can provide the designer with guidelines on the size of the array to be manufactured. If module placement is carried out for a fabricated array, area minimization frees up more cells for sample collection and preparation. We also introduce a simple measure, referred to as the fault tolerance index, to evaluate the fault tolerance capability of the microfluidic biochip; this measure is incorporated into the placement procedure. This procedure leads to small biochip area due to efficient utilization of dynamic reconfigurability, as well as high fault tolerance due to the efficient use of spare cells.

The organization of the remainder of the paper is as follows. In Section 2, we present an overview of digital microfluidics-based biochips. Section 3 discusses related prior work. In Section 4, we present a simulated annealing-based heuristic for module placement in dynamically reconfigurable biochips. Next, in Section 5, the reconfiguration technique is studied in more details, and the fault tolerance index, is defined. In Section 6, we incorporate the fault tolerance index in the placement procedure; we then use PCR to evaluate the enhanced placement procedure. Finally, conclusions are drawn in Section 7.

## 2. Background

The operation of digital microfluidics-based biochips is based on the principle of electrowetting actuation. Electrowetting refers to the modulation of the interfacial tension between a conductive fluid and a solid electrode by applying an electric field between them. The basic cell of a digital microfluidics-based biochip is shown in Figure 1(a). The droplet containing biochemical samples, and the filler medium, such as silicone oil, are sandwiched between two parallel glass plates. The bottom plate contains a patterned array of individually controllable electrodes, while the top plate is coated with a ground electrode. A hydrophobic dielectric insulator is added to the plates to decrease the wettability of the surface and to add capacitance between the droplet and the control electrode. By varying the electrical potential along a linear array of electrodes, nanoliter-volume droplets can transport along this line of electrodes. The velocity of the droplet (up to 20cm/s) can be controlled by adjusting the control voltage (0~90V). Microdroplets can therefore be moved freely to any location of a two-dimensional array without the need for pumps and valves. Figure 1(b) illustrates a fabricated microfluidic array [8].

Using a two-dimensional array, many common microfluidic operations for biomedical assays can be performed. For instance, the *mixing* operation is implemented by routing two droplets to the same location and then turning them around some pivot points. Note that these operations can be performed anywhere on the array during the operation of the biochip, whereas in continuous-flow systems they must operate in a specific permanently-etched micromixer or microchamber. This property of digital microfluidics-based biochips is referred to as dynamic reconfigurability, which we exploit here for high fault tolerance. The configurations of the microfluidic array are dynamically programmed into a microcontroller that controls the voltages of electrodes in the array. In this sense, the microfluidic modules (e.g., mixers or storage units) used during the operations can be viewed as reconfigurable virtual devices.

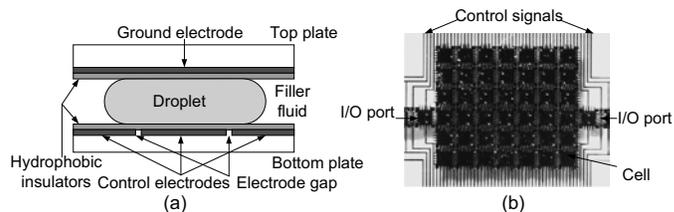

Figure 1: (a) Basic cell used in a digital microfluidics-based biochip; (b) A 2-D array for digital microfluidics.

## 3. Related Prior Work

Physical design automation for integrated circuits, especially module placement, is a mature topic [9]. Heuristics such as simulated annealing are extensively used for custom/macro cell placement; the placement problem is often formulated as 2-D rectangle packing [10]. Since these techniques do not consider reconfigurability, they are not directly applicable to programmable devices. Dynamically Reconfigurable FPGAs (DRFPGAs) have received much attention recently [11]. The partial reconfiguration offered by DRFPGAs is in many ways similar to the dynamic reconfigurability provided by digital microfluidics-based biochips. However, placement techniques reported thus far for DRFPGAs have only targeted the minimization of chip area [11]. Fault tolerance has not been considered in these placement techniques. Moreover, the programmability of DRFPGAs is limited by the well-defined roles of interconnect and logic blocks. Interconnect cannot be used for storing information, and logic blocks cannot be used for routing. In contrast, the digital microfluidics-based biochips offer significantly more programmability. The cells in the microfluidic array can be used for storage, functional operations, as well as for transporting fluid droplets.

As integrated circuits become denser, reliability emerges as a major challenge. Historically, reliability has been addressed through robust manufacturing processes. However, this approach does not address reliability issues associated with system design. Although microelectromechanical systems (MEMS) is a relatively young field compared to integrated circuits, reliability studies for MEMS have received considerable attention [12]. However, due to the significant differences in the actuation principles underlying between digital microfluidics and MEMS, these reliability enhancement techniques cannot be directly used for the design of microfluidics-based biochips.

While system-level physical design automation tools are now commonplace in integrated circuit design, no such efforts have been reported for digital microfluidics-based biochips. Some commercial computational fluidic dynamics (CFD) tools, such as CFD-ACE+ from CFD Research Corporation and FlumeCAD from Coventor, Inc. support 3-D simulation of fluidic transport. A recent release of CoventorWare from Coventor, Inc. includes microfluidic behavioral models to support system-level design. Unfortunately, this CAD tool is only able to deal with the continuous-flow systems, and it is therefore inadequate for the design of digital microfluidics-based biochips.

Recently, a fault classification and a unified test methodology for digital microfluidics-based biochips have been developed [13]. This cost-effective test methodology facilitates on-line testing, which allows fault testing and biochemical assays to run simultaneously on a microfluidics-based biochip [14].

## 4. Module Placement

Placement is one of the key physical design problems for digital microfluidics-based biochips. Based on the results obtained from architectural-level synthesis, i.e., a schedule of bioassay operation, a set of microfluidic modules, and the binding of bioassay operations to modules, placement determines the locations of each module on the microfluidic array in order to optimize some design metrics. Since digital microfluidics-based biochips enable dynamic reconfiguration of the microfluidic array during run-time, they allow the placement of different modules on the same location during different time intervals. Thus, the placement of modules on the microfluidic array can be modeled as a 3-D packing problem. Each microfluidic module is represented by a 3-D box, the base of which denotes the rectangular area of the module and the height denotes the time-span of its operation. The microfluidic biochip placement can now be viewed as the problem of packing these boxes to minimize the total base area, while avoiding overlaps.

Since placement follows architectural-level synthesis in the proposed synthesis flow, the starting times for each operation corresponding to a module, i.e., their positions in the time axis, are pre-determined. Therefore, the 3-D packing problem can be reduced to a modified 2-D placement problem. The horizontal cuts with the 3-D boxes correspond to the configurations of the microfluidic array at different point in time. For example, in Figure 3, the cut $t = t_1$ corresponds to a 2-D placement shown in Figure 2(b), and the cut $t = t_2$ corresponds to another configuration in Figure 2(c). The configurations of the microfluidic array during different time intervals can be combined together to form a modified 2-D placement shown in Figure 2(c). Note that the base of the 3-D box representing module $i$ should be placed on the cutting plane $t = S_i$, where $S_i$ is the starting time of module $i$'s operation determined by architectural-level synthesis. The modules can arbitrarily slide on these fixed cutting planes while avoiding overlap. Thus, instead of a 3-D packing problem, we only need to consider a modified 2-D placement consisting of several 2-D configurations in different time spans.

The module placement problem for electronic design is known to be NP-complete [7]. The microfluidic placement problem can also be shown by the method of restriction to be NP-complete. Consequently, heuristics are needed to solve the placement problem in a computationally efficient manner. Simulated annealing is a well-studied combinatorial optimization method, and it has been extensively used for traditional module placement problems [10]. An advantage of

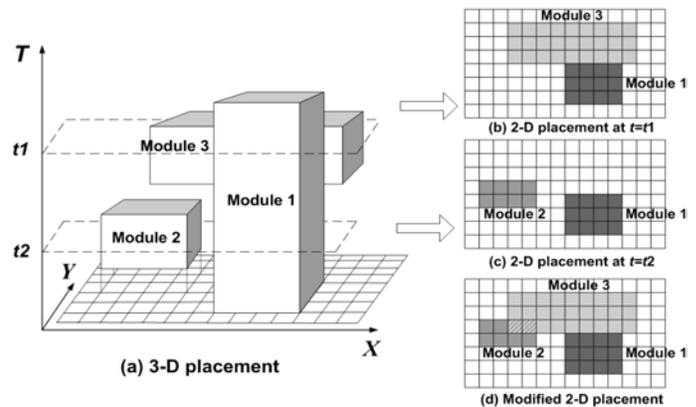

Figure 2: Reduction from 3_D placement to a modified 2-D placement.

**Procedure** PLACEMENT
/*Simulated Annealing-Based Module Placement */
1   $P = P_o$; /* Given initial placement */
2   $T = T_\infty$; /* Given initial temperature*/
3   $X = X_o$; /* Assign the annealing parameters */
4  **while** ("Stopping criterion" is not satisfied)
5     **for** i = 1: N   /* N is the number of iterations of the inner loop */
6       $P_{new}$ = **generate** (P); $\Delta C$ = **cost_metric**($P_{new}$)-**cost_metric**(P);
7       r = **random**(0, 1);  /* **random**(0, 1) is a function that returns a
8     pseudo-random number uniformly distributed on the interval [0, 1] */
9      **if**   $\Delta C < 0$ or $r < \exp(-\Delta C /T)$  { $P = P_{new}$ ;}
10    **end if**
11   **end for**   /* end of inner loop */
12     $T_{new} = \alpha \times T_{old}$; /* updating (cooling) temperature */
13  **end while**   /* end of annealing procedure */
14  output the optimal placement P.

Figure 3: Simulated annealing-based placement procedure.

simulated annealing is that it explores the configuration space of the optimization problem, while allowing *hill-climbing* moves, i.e., acceptance of new configurations that increase the cost. In this paper, we develop a simulated annealing-based algorithm to solve the placement problem for the digital microfluidics-based biochips. Instead of using a complicated problem-encoding scheme as in [15], our method directly applies the annealing procedure to the actual physical coordinates, sizes, and orientations of microfluidic modules. Since the direct approach cannot guarantee that each new placement is a feasible solution without any forbidden overlap, penalty for such forbidden overlaps is included in the cost function. The algorithm seeks to optimize the design metric while driving the overlap penalty to zero. The pseudocode for this heuristic approach is shown in Figure 3. Some important details of the algorithm are as follows.

a) *Initial placement*: It has been reported in the literature that the initial configuration has little impact on the final outcome of simulated annealing-based optimization [10]. Therefore, we apply a simple constructive approach to formulate the initial placement, as shown in Figure 4(a). In addition, during the annealing process, the modules are prevented from being placed outside the boundaries of the core area, as defined by Figure 4(a).

b) *Generation function*: New placements can be generated in several ways: (i) A single microfluidic module is randomly selected to be moved to a randomly-chosen location; (ii) A single module is randomly displaced to a new location and the orientation of this module is changed; (iii) A pair of modules are randomly selected for interchange; (iv) A pair of modules are interchanged in which at least one module has its orientation changed. During the annealing process, we assign the probability $p$ to the single-module displacement and $1-p$ to the two-module interchange. An effective ratio of $p/(1-p)$ is determined experimentally.

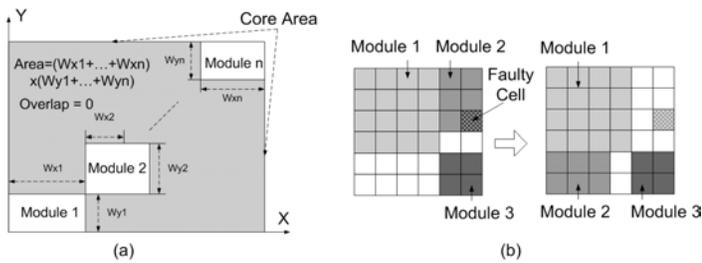

Figure 4: (a) Initial placement; (b) Example of partial reconfiguration.

c) *Controlling window for single-module displacement*: The displacement of a single module by a large distance leads to the large increase in the cost metric ($\Delta C > 0$). At low temperatures during the annealing process, only the new generations with $\Delta C \leq 0$ have a reasonable chance of being accepted. This increases the probability that the displacements over large distances are rejected. We apply a controlling window to discourage long-distance displacements at low temperatures. As the temperature approaches zero, the span of the controlling window reaches its minimum value; this condition is used as the stopping criterion for simulated annealing.

d) *Annealing scheme*: Most annealing parameters are experimentally determined. These include the following: (i) The temperature is modulated as $T_{new} = \alpha \times T_{old}$, where $\alpha = 0.9$; (ii) The number of iterations of the inner loop for a given value of $T$ is determined using the relationship $N = Na \times Nm$, where $Na = 400$ and $Nm$ is the number of the modules; (ii) The initial temperature $T_\infty$ is chosen to ensure that almost every new placement can be accepted, here $T_\infty = 10000$.

e) *Cost metrics*: Cost metrics are used to mathematically represent the optimization goals of the placement problem. We consider the area of the array and the degree of fault tolerance as cost metrics.

## 5. Fault Tolerance and Reconfiguration

In this section, we investigate dynamic and partial reconfiguration to avoid a faulty cell in the microfluidic array. Based on this reconfiguration technique, a simple numerical measure, termed fault tolerance index, is defined to estimate the fault tolerance capability of the biochip. We also present an efficient algorithm to determine the fault tolerance index of a biochip configuration based on the notion of maximal-empty rectangles.

### 5.1 Partial Reconfiguration

A digital microfluidics-based biochip can be viewed as a dynamically reconfigurable system. If a cell becomes faulty during the operation of the biochip, detected using the technique described in [13], the microfluidic module containing this cell can easily be relocated to another part of the microfluidic array by changing the control voltages applied to the corresponding electrodes. An example of partial reconfiguration is shown in Figure 4(b). Fault-free unused cells in the array are utilized to accommodate the faulty module. Hence, the configuration of the microfluidic array, i.e., the placement of the microfluidic modules, influences the fault tolerance capability of the biochip. Moreover, since partial reconfiguration only targets the module containing the faulty cell and leaves other aspects of the microfluidic configuration unchanged, a fast heuristic algorithm can be used to find a new location for this module. Therefore, partial reconfiguration is suitable for dynamic on-line reconfiguration during field operation of the microfluidic biochip.

### 5.2 Fault Tolerance Index

In order to facilitate partial reconfiguration and incorporate fault tolerance in the simulated annealing-based placement procedure, we need to evaluate the fault tolerance capability of the microfluidic biochip.

We consider the reconfiguration problem for a single failing cell in the microfluidic array. The single fault assumption is valid when testing and reconfiguration are carried out frequently. We also assume that every cell has the same failure probability. Since microfluidic biochips have not yet been manufactured in large numbers, failure data or statistical models are not readily available, and the assumption of uniform failure probability is reasonable. The failure model can be easily updated when statistical failure data becomes available.

We use a 2-D coordinate system to refer to the cells in the microfluidic array. The bottom-left cell is referred to as (1, 1) and the top-right cell in an $m \times n$ array is referred to as ($m$, $n$). For an $m \times n$ microfluidic array, assume that an arbitrary cell ($i$, $j$) is faulty. If this cell is contained in a module for a given microfluidic configuration $C$, we attempt to apply partial reconfiguration to relocate this module to avoid the faulty cell. If this reconfiguration succeeds, i.e., we find an adequate number of contiguous cells to accommodate this module, or cell ($i$, $j$) is not used by any module, we deem this cell to be *C-covered* for this configuration. Otherwise, cell ($i$, $j$) is not *C*-covered. For an array with $k$ *C*-covered cells, we define the *fault tolerance index* (FTI) as follows: FTI = $k/(m \times n)$.

Noted that FTI lies between 0 and 1. It increases if there are more *C*-covered cells in the array. If FTI is 1, it implies that when any single cell in the array is faulty, this microfluidic configuration can be used by applying partial reconfiguration to bypass the faulty cell. On the other hand, if FTI is 0, the biochip cannot be reconfigured if any arbitrary cell becomes faulty. This is the worst case that needs to be avoided.

In order to determine if a cell is *C*-covered for configuration *C*, we use an efficient procedure based on the notion of maximal-empty rectangles. The details of this procedure are described below.

### 5.3 Fast algorithm to determine FTI

Our goal is to find maximal-empty rectangles in the microfluidic array, and then check if these rectangles can accommodate the faulty module. A *maximal empty rectangle* is defined as an empty rectangle (a set of unused cells) that cannot be completely covered by any other empty rectangles. If a maximal-empty rectangle can accommodate the faulty module, this module can be relocated to the empty rectangle to avoid the faulty cell. If no such maximal-empty rectangle exists, partial reconfiguration is deemed to have failed. We then conclude that the corresponding faulty is not *C*-covered.

An encoding method is first used to facilitate the implementation of this algorithm. If a module contains a faulty cell, this module is temporarily removed from the placement. Next the configuration of the microfluidic array is modeled by a matrix consisting of 0s and 1s. The faulty cell and all cells contained in the currently operational modules are represented by 1s; all unused cells are represented by 0s.

In order to find all maximal-empty rectangles rapidly, a data structure, referred to as the staircase [16], is employed in the algorithm. A *staircase*($x$, $y$) is defined as the collection of all overlapping empty rectangles with ($x$, $y$) as their bottom-right corner.

The data structure *staircase*($x$, $y$) help to determine all maximal-empty rectangles that lie entirely within the *staircase*($x$, $y$) and whose bottom-right corner is ($x$, $y$). The algorithm traverses the matrix left-to-right and top-to-bottom, creating a staircase for every cell in the matrix. Next, based on the knowledge of staircases, all maximal-empty rectangles are determined. The details underlying the construction of staircases and the generation of the maximal-empty rectangles from staircases are described in [16].

## 6. Experimental Evaluation: PCR

Polymerase chain reaction (PCR) is one of the most common techniques for DNA analysis [17]. It is used for rapid enzymatic amplification of specific DNA fragments. PCR can amplify genomic DNA exponentially using temperature cycles. Recently, the feasibility of performing droplet-based PCR on digital microfluidics-based biochips has been successfully demonstrated [2]. In this section, we use the mixing stage of PCR as an example to evaluate the simulated annealing-based placement algorithm that facilitates fault tolerance. Its assay protocol can be modeled by a sequencing graph [17], as shown in Figure 5.

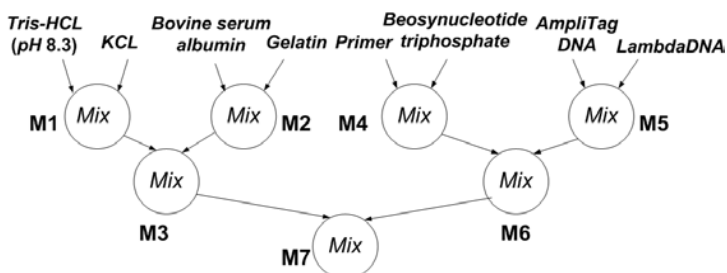

Figure 5: Sequencing graph for the mixing stage of PCR.

Based on this graph model, architectural-level synthesis can be used to carry out both resource binding and scheduling. Let the resource binding be as shown in Table 1. Note that the module generated here has a segregation region wrapped around the functional region, which not only isolates the functional region from its neighbors but also provides a communication path for droplet movement. The data for the operation times associated with the different modules are obtained from real-life experiments [18]. A schedule for the functional operations and module usage is shown in Figure 6.

Table 1: Resource binding in PCR.

| Operation | Hardware* | Module | Mixing time |
|---|---|---|---|
| M1 | 2x2 electrode array | 4x4 cells | 10s |
| M2 | 4-electrode linear array | 3x6 cells | 5s |
| M3 | 2x3 electrode array | 4x5 cells | 6s |
| M4 | 4-electrode linear array | 3x6 cells | 5s |
| M5 | 4-electrode linear array | 3x6 cells | 5s |
| M6 | 2x2 electrode array | 4x4 cells | 10s |
| M7 | 2x4 electrode array | 4x6 cells | 3s |

*: Electrode pitch: 1.5 mm; Gap height: 600 μm

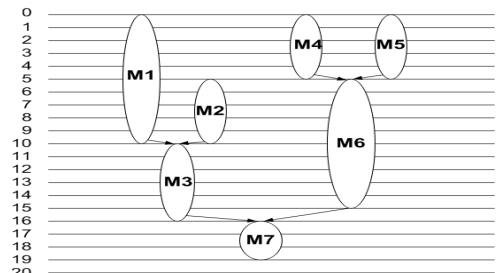

Figure 6: Schedule highlighting the usage of microfluidic modules.

### 6.1 Enhanced module placement algorithm

We use a simple greedy algorithm as a baseline for assessing the quality of the proposed placement method. Modules are first sorted in the descending order based on their areas. In each step, the module with the largest area among the unplaced ones is selected and placed at an available bottom-left corner of the array. The total area of the placement generated is 189mm$^2$, i.e., it consists of 84 cells, where the pitch of each cell is 1.5mm.

Next we apply the placement procedure of Section 4 to this example. First, we consider the minimization of the array area as the only cost metric. The placement generated by the simulated annealing procedure is shown in Figure 7. Its total area is 141.75mm$^2$ (63 cells), which is 25% less compared to the baseline. The computation takes 5 minutes of CPU time on a 1.0 GHz Pentium-III PC with 256 MB of RAM. Note that some microfluidic modules, e.g., Modules 1 and 3, can use the same cells (via dynamic reconfiguration) when their time-spans do not overlap.

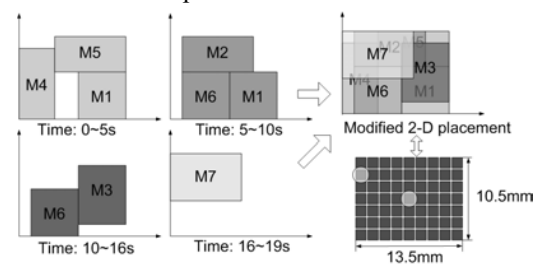

Figure 7: Placement obtained from the simulated annealing-based procedure (7x9=63 cells).

Due to the efficient utilization of dynamic reconfigurability, the algorithm leads to a highly compact placement. However, the placement with the minimum array area does not provide adequate fault tolerance. We determine the FTI of the placement shown in Figure 7 using the fast algorithm described in Section 5.3. (The calculation of FTI takes only 1.7 seconds of CPU time.) The FTI of this design is only 0.1270, which implies that only 8 cells in this 7×9 array are *C*-covered. A microfluidics-based biochip with such a low degree of fault tolerance is not suitable for critical DNA analysis.

The goal of the enhanced placement algorithm is to maximize FTI, while keeping the total biochip area small. FTI and area are conflicting criteria, because high FTI often requires a larger biochip area. In our multi-objective placement problem, a solution is a 2-tuple (area, FTI) resulting from a feasible placement of microfluidic modules.

Weighting is a commonly used method for multi-objective optimization. A weight is assigned to each objective according to its relative importance. Next, the different objectives are combined into a single objective using a weighted sum. The solution with the lowest weighted sum is selected. In our problem, weights $\alpha$ and $\beta$ are assigned to the criteria of area and FTI, respectively. We set $\alpha$ to 1 and adjusted $\beta$ according to the degree of importance of fault tolerance. The solution with the lowest value of the metric ($\alpha \times$area$-\beta \times$fault-tolerance number) was considered to be an acceptable solution.

We used a two-stage simulated annealing-based algorithm. In the first stage, a fault-oblivious simulated annealing-based algorithm is used to obtain a placement with the smallest area. Starting from this intermediate configuration, the second stage uses low-temperature simulated annealing (*LTSA*) to refine the placement in order to enhance fault tolerance. The FTI measure is included in the cost function, while the total area is kept as small as possible. In addition, during *LTSA*, only single module displacement is performed.

The two-stage method takes 20 minutes of CPU time. The solution obtained requires an area of 173.25mm$^2$, and yields an FTI of 0.8052; see Figure 8. In comparison to the previous placement with less area, this solution leads to an increase of 534% in FTI, while increasing the area by only 22.2%. This is clearly a more desirable placement for the safety-critical PCR assay.

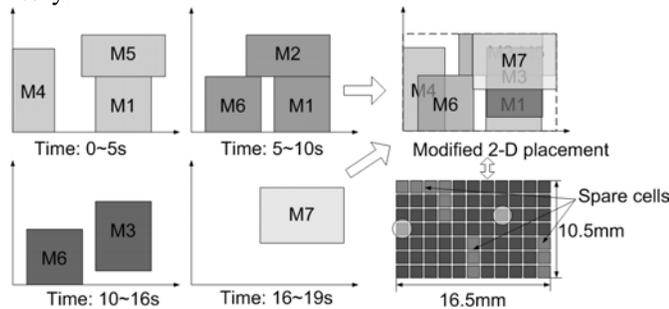

Figure 8: Placement obtained from the enhanced module placement algorithm: 7x11 = 77 cells, and FTI = 0.8052.

### 6.3 Effect of $\beta$

The parameter $\beta$ represents the importance of fault tolerance to the biochip designer. If fault tolerance is critical, e.g., for implantable microfluidic drug-dosing systems, a relatively large value of $\beta$ can be used to increase FTI. On the other hand, if fault tolerance is less important, e.g. for disposable carry-home glucose detectors for one-time use, a relatively small value of $\beta$ can be used, thereby reducing area and product cost.

In the final set of experiments, we varied $\beta$ to investigate the relationship between chip area and FTI (Table 2). With an area of 222.75mm$^2$, we can ensure that the system can always tolerate one faulty cell.

Table 2: Solutions for different value of $\beta$.

| $\beta$ | 10 | 20 | 30 | 40 | 50 | 60 |
|---|---|---|---|---|---|---|
| Area (mm$^2$) | 141.75 | 157.5 | 173.25 | 189.0 | 204.75 | 222.75 |
| *FTI* | 0.2857 | 0.7143 | 0.8052 | 0.8571 | 0.9780 | 1.0 |

## 7. Conclusions

We have presented a simulated annealing-based technique for module placement in microfluidic biochips. The placement criteria include chip area as well as fault tolerance; the latter allows a microfluidic module to be relocated elsewhere in the system when a single cell is detected to be faulty. The placement problem accounts for dynamic reconfigurability of droplet-based microfluidics, whereby groups of cells can be reconfigured to change their functionality during the concurrent execution of a set of bioassays. We have presented simulation results for a case study involving the polymerase chain reaction. This work is expected to facilitate the automated design of biochips, especially since their complexity is expected to grow steadily as they are increasingly used for clinical diagnosis, DNA sequencing, and other laboratory procedures involving molecular biology.